\documentclass{article}
\usepackage{bm}
\usepackage{amsmath}
\usepackage{graphicx}
\usepackage{authblk}
\usepackage[numbers,sort&compress]{natbib}
\title{Fragmentation instability in aggregating systems}

\author[1,*]{Arturo Berrones-Santos}
\author[2]{Luis Benavides-V\'azquez}
\author[1]{Elisa Schaeffer}
\author[3]{Javier Almaguer}

\affil[1]{\small Posgrado en Ingenier\' \i a de Sistemas, Facultad de de Ingenier\' \i a Mec\'anica y El\'ectrica, Universidad Aut\'onoma de Nuevo Le\'on, San Nicol\'as de los Garza, N.L., M\'exico}
\affil[2]{\small School of Engineering and Sciences, Tecnologico de Monterrey, Monterrey, N.L., Mexico}
\affil[3]{\small Doctorado en Ciencias con Orientaci\'on en Matem\'aticas, Facultad de Ciencas F\'isico Matem\'aticas, Universidad Aut\'onoma de Nuevo Le\'on, San Nicol\'as de los Garza, N.L., M\'exico}
\affil[*]{\small \textbf{Corresponding author}: arturo.berronessn@uanl.edu.mx}

\begin{document}
\maketitle

\begin{abstract}
The inclusion of a fragmentation mechanism in 
population balance equations introduces complex interactions that
make the analytical or even computational
treatment much more
difficult than for the pure aggregation case.
This is specially true when 
variable sized fragments are allowed, because 
of the exponential growth in fragments size combinations
with the number of monomers in the exchanges.
In this contribution
we present a new model that incorporates an
instability threshold in the clusters, which induces
arbitrary losses or gains of particles by fracture with
a substantial simplification of the combinatorics of the process.
The model exhibits two different regimes.
\end{abstract}

\section{Introduction}
\label{Introduction}

Despite that aggregation processes have been under study since the $1910$'s after
the seminal work by Smoluchowski \cite{Smoluchowski}, the understanding of the phenomena that emerge from these 
processes is far to be complete. For instance, only very recently has been formally proved
that the gelation without fracture observed not only in colloids \cite{poon}
but also in the form of a ``giant component''
present in such disparate systems like stars \cite{allen} or social networks \cite{kumar,barabasi}
is indeed a thermodynamic phase transition \cite{matsoukas}. 
The inclusion of fragmentation mechanisms only increases the number of interesting behaviors 
like the emergence of stationary states with unusual statistical properties \cite{majumdar,vigil}, 
which are relevant for applications like biological
networks \cite{bressloff,pakdaman}, econophysics \cite{chakrabarti} or
chemical and physical systems out of equilibrium \cite{almaguer}.
The classical framework to deal with aggregation-fragmentation processes is given by the
generalized Smoluchowski rate equation \cite{vigil},
\begin{eqnarray} \label{af}
\frac{d\rho_n}{dt} = \frac{1}{2} \sum_{i+j=n} \left[ K_{i,j} \rho_i \rho_j - F_{i,j} \rho_{i+j} \right ] \\ \nonumber
- \sum_{j = 1} ^ {\infty} \left [ K_{n,j} \rho_{n} \rho_{j} - F_{n,j} \rho_{n+j} \right ].
\end{eqnarray}
Equation (\ref{af}) describes the evolution of the density of clusters with size $n$, $\rho_n$, in terms of the
aggregation kernel $K_{i,j}$ (which gives the aggregation reaction rate of an $i$-mer with a $j$-mer) and the 
fragmentation kernel $F_{i,j}$ (models the break-up of a ($i+j$)-mer into an $i$-mer and a $j$-mer).
It should be remarked that Eq.~\eqref{af} is a mean field description, which gives no information regarding spatial 
fluctuations.
The study of the mathematical properties of Eq.~\eqref{af} is an active research area \cite{banasiak,lamb,bansiak2},
however the combinatorics introduced by the fragmentation terms make the analytical treatment very
difficult even with simple kernel structures. From a computational standpoint, the presence of fragmentation
makes the numerical solution of Eq.~\eqref{af} also in general intractable, because the evident 
exponential computing times unless the fracture exchanges are in some way limited. 
The most studied special case has been the situation with single particle fractures, like in the ``one-chip'' or 
in the Becker-D\"oring models \cite{majumdar,vigil,wattis}.
Very recently it has been also put forward a quite
general setup in which the fragmentation is complete, that is, it results in a full
decomposition into monomers
\cite{bodrova}. Under this assumption, a thorough analytical and numerical characterization of
a variety of aggregation mechanisms is given in \cite{bodrova}.

In the present work we introduce a new model that allows 
fragments with an
arbitrary number of monomers but with simple combinatorics.
The model is inspired in processes in which ``large'' clusters behave in a different way than ``small'' clusters.
Our original motivation was the jar test apparatus for wastewater treatment 
reported in \cite{almaguer}, for which some of the authors of the present
contribution developed a cellular automaton model. A mean field
description for processes with the same general microscopic 
properties is the purpose of the present contribution.
In the jar test apparatus,
a transition from gel states to stationary particle size 
distributions in which very large clusters are dominant is observed. The transition is
induced by instabilities in the large clusters, controlled by an external energy input.
Similar situations may also 
arise for instance in socio-economic systems, where large organizations can fracture into smaller ones
but also can acquire parts of other large organizations or entirely absorb small ones, while small organizations are able to
grow only by aggregation of simple units of capital or shares.
In these kind of instances, there is a natural asymmetry by which large clusters behave like hubs for
small clusters but also display a tipping point after which they
become intrinsically unstable. 
There is a number of applications in which this emergence of asymmetry 
after a threshold is reported \cite{bak,faybishenko,lomnitz,markovic}.
In first instance our model can be adapted to these situations and 
in consequence an open system
in which mass conservation can be violated is tacitly assumed.
The closed system version of the model will be reported elsewhere.
The statistical properties of the stationary regimes can be derived from a 
characteristic function in a number of general cases
(for a discussion of the use of characteristic functions in statistical physics,
specially the technique based on the Z-function, which is the one followed by us 
in our analytical description below, 
see for instance \cite{krapivsky}).

\section{Instability threshold aggregation-fragmentation model}
\label{model}

The basic feature of our model is the introduction of a global instability threshold, $\bar{h}$. This can be interpreted as a
physical constant that acts on all of the clusters in the system, similar to those used in models for sandpiles \cite{bak}, 
earthquakes \cite{lomnitz,markovic} and fracture of materials \cite{faybishenko}. To our knowledge, our model is the first that adds such a mechanism in 
aggregation-fragmentation processes modeled by Smoluchowski-type rate equations. 
Our basic concept is that clusters with a number of particles above $\bar{h}$ became unstable and can fracture with a 
constant probability. A fracture gives rise to two clusters, one of which is barely stable with size $\bar{h}$. From a microscopic
level, one may say that the original cluster has transferred its ``excess'' particles.
Fragments from an originally large cluster can be absorbed by other large clusters
above $\bar{h}$. This introduces an asymmetry between clusters below and above
the instability threshold, which implies an open system 
in which there can be a flux
of mass out of the system to a gel state.
Interestingly, as shown below,
stationary states can still emerge despite this fact.
The gel regime on the other hand, is attained in a finite time.

The inclusion of the parameter $\bar{h}$ permits a radical simplification of the 
fragmentation terms of the Smoluchowski-type model 
without losing the exchanges resulting in large fragments present in the general case, at the cost of the need for two regimes,
\begin{eqnarray} \label{smol}
n = i + j > \bar{h}: \\ \nonumber
\frac{d\rho_n}{dt} = \frac{1}{2} \sum_{i+j=n} K_{i,j} \rho_i \rho_j - \sum_{j = 1} ^ {\infty} K_{n,j} \rho_{n} \rho_{j}
 - \frac{F}{2} \rho_{n} + F \rho_{n+\bar{h}}. \\ \nonumber
n = i + j \leq \bar{h}: \\ \nonumber
\frac{d\rho_n}{dt} = \frac{1}{2} \sum_{i+j=n} K_{i,j} \rho_i \rho_j - \sum_{j = 1} ^ {\infty} K_{n,j} \rho_{n} \rho_{j}.
\end{eqnarray}
The possibly exponentially growing combinatorics obviously persists in the aggregation terms, but these are
tractable for some general kernels $K_{i,j}$ due to the separability in the $\rho$'s. 

\section{Related recent literature}
A number of advances to tackle generalizations of population balance
models beyond constant kernels and with spontaneous fragmentation have been 
very recently put forward \cite{bodrova,kaur,brilliantov,timokhin}. 
The main difference of the setup given by the expressions
(\ref{smol}) with respect to these other recent approaches
is the discontinuous nature of the spontaneous fragmentation probability
induced by an instability threshold, which can lead to states 
with a population dominated by very large fragments or can go to a gel phase.
For instance, in the already mentioned work \cite{bodrova}, analytical solutions
under collisional kernels are given, supporting aggregation and spontaneous and 
collisional fragmentation. This model clearly represents a substantial step beyond
the classical theoretical results under simple aggregation and fragmentation mechanisms.
However, the fragmentation is subjected to a regularity condition that  
limits the process to be ``complete'', that is, it results in
a division in the basic constituent monomers. This is suitable for cases in 
which small monomers or debris dominate the particle size distribution.
An application to particle size distributions in Saturn's
rings is given of this collisional model is given in \cite{brilliantov}  and 
in \cite{timokhin} a numerical approach for such type of models with collisional 
fragmentation is given.
In \cite{kaur},
by writing the population rate equation in integral form, 
the authors find exact solutions by the so called homotopy method. 
The approach is quite general and admits spontaneous fragmentation by assuming 
continuous fragmentation rates. It would be an interesting 
future research direction
to investigate the possible implementation of this approach to our
discontinuous setup.

From a computational standpoint, fast algorithms for the direct numerical solution of 
the rate equations have been recently proposed. In particular 
\cite{osinsky} is capable to deal
with very general kernel structures by a low-rank transformation 
at the cost of ignoring the tails of the particle size distribution. 
This at first instance seems not well suited for our model in which
as already mentioned, the existence of states with large particle sizes is of interest.
It may however be fruitfully adapted to the study of some of our
stationary regimes. In its standard formulation this fast algorithm (and other 
recent proposals like \cite{smirnov}) does not consider fragmentation, so generalizations 
are in order to explore our instability model by direct numerical solution.
We have proposed a cellular automaton like an at first instance best suited numerical
framework for our model. 
Our original motivation for the associated cellular automaton model has been a waste water treatment
experiment in a jar test apparatus in which by the injection of external energy, agitation favours the large
flock formation, but when the flocks exceed a certain threshold, instabilities emerge that can cause fracture
\cite{almaguer}. Other application that in our opinion is worth to be explored in the future is in
the fragmentation of social networks. There is an increasing body of evidence
that in networks
of individuals or of social organizations, the fragmentation is dominated by tipping points and
not by the continuous fragmentation rates usually studied in population balance approaches. 
See for instance \cite{minh} for discussions of fragmentation in social contexts. 
Community formation processes have been extensively studied by complex networks models \cite{krapivsky,antal,barabasi},
but the study of models for community fragmentation is still scarce \cite{bome}.
Our Smoluchowski-type setup can be a useful mean field description of these 
social networks processes.

\section{Analysis under different aggregation kernels}
\subsection{Constant aggregation rate}

With constant aggregation kernel, the model reduces to
\begin{eqnarray} \label{smol2}
n = i + j > \bar{h}: \\ \nonumber
\frac{d\rho_n}{dt} = \frac{1}{2} K \sum_{j=1}^{n-1} \rho_j \rho_{n-j} - K \sum_{j = 1} ^ {\infty} \rho_{n} \rho_{j}
 - \frac{F}{2} \rho_{n} + F \rho_{n+\bar{h}}. \\ \nonumber
n = i + j \leq \bar{h}: \\ \nonumber
\frac{d\rho_n}{dt} = \frac{1}{2} K \sum_{j=1}^{n-1} \rho_j \rho_{n-j} - K \sum_{j = 1} ^ {\infty} \rho_{n} \rho_{j}
\end{eqnarray}
The Z-transform (characteristic function) $\hat{\rho}(z) = \sum_{n=1}^{\infty} \rho_{n} z^{-n}$ of the model (\ref{smol2}) 
reads
\begin{eqnarray} \label{ricatti}
\frac{d\hat{\rho}(z,t)}{dt} = \frac{K}{2} \hat{\rho}^{2}(z,t) - \hat{\rho}(z,t) \left[ K\hat{\rho}(1,t) + \frac{F}{2}(1-2z^{\bar{h}}) \right ] 
-F \sum_{j=1}^{\bar{h} - 1} \rho_j z^{\bar{h} - j} 
\end{eqnarray}
which admits stationary solutions,
\begin{eqnarray} \label{zstat}
\hat{\rho}(z) = \hat{\rho}(1) + \frac{F}{2K} (1-2z^{\bar{h}})
\pm \frac{1}{2}\sqrt{\left[ 2\hat{\rho}(1) + \frac{F}{K} (1-2z^{\bar{h}}) \right]^2 
+ 8\left(\frac{F}{K} \right) \sum_{j=1}^{\bar{h} - 1} \rho_j z^{\bar{h} - j}},
\end{eqnarray}
where
\begin{eqnarray} \label{zonesol}
\hat{\rho}(1) = \frac{F}{2K} \pm \frac{1}{2}\sqrt{\left( \frac{F}{K} \right ) ^2 - 8 \left( \frac{F}{K} \right ) \sum_{j=1}^{\bar{h} - 1} \rho_j }
\end{eqnarray}
is the total number of clusters in the system. 
In the stationary regime, $1 / \hat{\rho}(1)$ is a normalization factor and the moments of the stationary
cluster size distribution follow from the derivatives of the characteristic function.
For instance, $\left < n \right > = \frac{-1}{\hat{\rho}(1)} \frac{d\hat{\rho}}{dz} |_{z=1}$ and 
$\left < n(n+1) \right > = \frac{1}{\hat{\rho}(1)} \frac{d^{2}\hat{\rho}}{dz^{2}} |_{z=1}$.
By taking derivatives in equation (\ref{ricatti}), a closed system for the first two moments in terms of the
normalization factor is obtained,
\begin{eqnarray}
-K\hat{\rho}(1) \left < n \right > + \left < n \right > \left [ K \hat{\rho}(1) - \frac{F}{2} \right ]
+ F\bar{h} - \frac{F}{\hat{\rho}(1)} \sum_{j=1}^{\bar{h} - 1} (\bar{h} - j) \rho_j = 0 \\ \nonumber
K\hat{\rho}(1) \left < n(n+1) \right > + K\hat{\rho}(1) \left < n \right >^2 - 2F\bar{h}\left < n \right > \\ \nonumber
- \left < n(n+1) \right > \left [ K \hat{\rho}(1) - \frac{F}{2} \right ] + \left [ F\bar{h} (\bar{h} - 1) \right ] \\ \nonumber
- \frac{F}{\hat{\rho}(1)} \sum_{j=1}^{\bar{h} - 1} (\bar{h} - j) (\bar{h} - j - 1)\rho_j = 0,
\end{eqnarray}
from which,
\begin{eqnarray} \label{moments1}
\frac{1}{2} \left < n \right > = \bar{h} - \frac{1}{\hat{\rho}(1)} \sum_{j=1}^{\bar{h} - 1} (\bar{h} - j) \rho_j, \\ \nonumber
\frac{1}{2} \left < n(n+1) \right > = 2 \left < n \right > \bar{h} - \frac{K}{F}\hat{\rho}(1) \left < n \right >^2
- \bar{h} (\bar{h} - 1) + \frac{1}{\hat{\rho}(1)} \sum_{j=1}^{\bar{h} - 1} (\bar{h} - j) (\bar{h} - j - 1)\rho_j .
\end{eqnarray}
By introducing the definitions, $\alpha \equiv \frac{1}{\hat{\rho}(1)} \sum_{j=1}^{\bar{h} - 1} \rho_j$,
$\beta \equiv \frac{1}{\hat{\rho}(1)} \sum_{j=1}^{\bar{h} - 1} j \rho_j $ and 
$\gamma \equiv \frac{1}{\hat{\rho}(1)} \sum_{j=1}^{\bar{h} - 1} j^2 \rho_j$ follows that,
\begin{eqnarray} \label{moments2}
\left < n \right > &=& \frac{\bar{h}(1 - \alpha)}{\frac{1}{2} - \beta}, \\ \nonumber
\left < n^2 \right > (1 - 2\gamma) &=& \left < n \right > (4\bar{h} - 1) - \frac{2K}{F}\hat{\rho}(1) \left < n \right >^2
- 2\bar{h} (\bar{h} - 1) \\ \nonumber
&+& 2(\alpha \bar{h}^2 - 2\beta \bar{h} \left < n \right > - \alpha  \bar{h} + \beta \left < n \right >),
\end{eqnarray}
so the first two moments of the stationary distribution are well defined under the conditions 
$0 < \alpha < 1$, $0 < \beta < 1/2$, $\gamma > 1/2$. The mean grows linearly with $\bar{h}$
and the variance is parabolic in $\bar{h}$.
At small but non zero $(F/K)$, mean and variance are large and diverge in the
limit $(F/K) \to 0$
while $\hat{\rho}(1) \to 0$, which corresponds to the limit $\beta \to 1/2$.
This kind of stationary state is attained in finite time.
To see this, consider $(F/K) \ll 1$. From Eq.(\ref{zonesol}) follows that,
\begin{eqnarray}
\frac{F}{K} = \frac{1}{1-\frac{2}{\alpha}\sum_{j=1}^{\bar{h} - 1} \rho_j}
\end{eqnarray}
Note also that at $(F/K) \ll 1$ the sum $\sum_{j=1}^{\bar{h} - 1} \rho_j \ll \hat{\rho}(1)$, so
\begin{eqnarray}
\frac{d\hat{\rho}(1,t)}{dt} = - \frac{K}{2} \hat{\rho}^{2}(1,t) + \frac{F}{2} \hat{\rho}(1,t),
\end{eqnarray}
which leads to
\begin{eqnarray}
\hat{\rho}(1,t) = \frac{F}{K[1-e^{-(\frac{F}{2})t}]}
\end{eqnarray}
Therefore, the system displays a stationary state with
large mean and variance at $(F/K) \ll 1$ and above a characteristic
time $t_c \approx \frac{2}{F}$.
This behavior is absent in the pure aggregation model
with constant kernel, where a gelation transition occurs at infinite time.

\subsection{Additive and multiplicative aggregation rate}

For aggregation kernels of the form $K_{i,j} = K(i+j) / 2$ and $K_{i,j} = Kij$, where $K$ is some positive constant,
the resulting system of equations for the 
moments is not closed, but however some important features of the stationary states can be derived.
 
In the additive case, 
\begin{eqnarray} \label{additive}
n = i + j > \bar{h}: \\ \nonumber
\frac{\partial \rho_n}{\partial t} = \frac{Kn}{4} \sum_{j=1}^{n-1} \rho_j \rho_{n-j} - \frac{Kn\rho_{n}}{2} \sum_{j = 1} ^ {\infty} \rho_{j}
- \frac{K\rho_{n}}{2} \sum_{j = 1} ^ {\infty} j \rho_{j}
- \frac{F}{2} \rho_{n} + F \rho_{n+\bar{h}}. \\ \nonumber
n = i + j \leq \bar{h}: \\ \nonumber
\frac{\partial \rho_n}{\partial t} = \frac{K}{4} \sum_{i+j=n} (i+j) \rho_i \rho_j - \frac{K}{2} \sum_{j = 1} ^ {\infty} (n+j) \rho_{n} \rho_{j}.
\end{eqnarray}
and the Z-transform of the model reads,
\begin{eqnarray} \label{add}
\frac{\partial \hat{\rho}(z,t)}{\partial t} = - \frac{K}{2} z \hat{\rho}(z,t) \frac{\partial \hat{\rho}(z,t)}{\partial z} 
+ \frac{K}{2} z \hat{\rho}(1,t) \frac{\partial \hat{\rho}(z,t)}{\partial z} 
- \frac{K}{2} M_1(t) \hat{\rho}(z,t) \\ \nonumber
- \frac{F}{2}(1-2z^{\bar{h}}) \hat{\rho}(z,t)
-F \sum_{j=1}^{\bar{h} - 1} \rho_j z^{\bar{h} - j} ,
\end{eqnarray}
where $M_1(t) \equiv \sum_{j = 1} ^ {\infty} j \rho_{j}$. If the definition for $\left < n \right >$ is introduced in
(\ref{add}), it turns out that there exists a stationary state in which
\begin{eqnarray}
\left < n \right > = \frac{F}{K} \left [ 1 - 2 \left ( \frac{\sum_{j=1}^{\bar{h} - 1} \rho_j}{\hat{\rho}(1)} \right ) \right ],
\end{eqnarray}
given the condition $\hat{\rho}(1) > 2 \sum_{j=1}^{\bar{h} - 1} \rho_j$. 

The product aggregation kernel $K_{i,j} = Kij$ leads to
\begin{eqnarray} \label{product}
n = i + j > \bar{h}: \\ \nonumber
\frac{d\rho_n}{dt} = \frac{1}{2} \sum_{i+j=n} ij \rho_i \rho_j - \sum_{j = 1} ^ {\infty} nj \rho_{n} \rho_{j}
 - \frac{F}{2} \rho_{n} + F \rho_{n+\bar{h}}. \\ \nonumber
n = i + j \leq \bar{h}: \\ \nonumber
\frac{d\rho_n}{dt} = \frac{1}{2} \sum_{i+j=n} ij \rho_i \rho_j - \sum_{j = 1} ^ {\infty} nj \rho_{n} \rho_{j} ,
\end{eqnarray}
with the associated characteristic function,
\begin{eqnarray} \label{mult}
\frac{\partial \hat{\rho}(z,t)}{\partial t} = \frac{K}{2} z^2 \left [ \frac{\partial \hat{\rho}(z,t)}{\partial z} \right ]^2
+K z M_1(t) \frac{\partial \hat{\rho}(z,t)}{\partial z} \\ \nonumber
- \frac{F}{2}(1-2z^{\bar{h}}) \hat{\rho}(z,t)
-F \sum_{j=1}^{\bar{h} - 1} \rho_j z^{\bar{h} - j}
\end{eqnarray}
The pure aggregation case ($F=0$) has been widely studied and display a number of interesting behaviors \cite{wattis}.
In the presence of fragmentation, stationary states emerge. To procede substitute the definition for $\left < n \right >$ into
equation (\ref{mult}), giving
\begin{eqnarray} \label{multsol}
\left < n \right > = 
\sqrt{\frac{F}{K} \frac{\left [ \sum_{j=1}^{\bar{h} - 1} \rho_j - \frac{1}{2}\hat{\rho}(1) \right ]}
{\hat{\rho}(1)\left [ \frac{1}{2}\hat{\rho}(1) - 1 \right ]}}
\end{eqnarray}
so there is a stationary state in which all the mass is concentrated in a finite number of clusters. The total 
number of clusters is in the interval $2 \leq \hat{\rho}(1) \leq 2 (\bar{h} - 1)$.

\section{Microscopic justification of the instability threshold: a cellular automaton model}

A microscopic counterpart of the constant kernels version of our model can be given in terms of the
following cellular automaton (which generalizes a model introduced by Almaguer et
al.\ \cite{almaguer}): consider
a $n$-cell $d$-dimensional wrap lattice (for example, in a $1$-d automaton, the first cell is adjacent with the last cell). 
The cell coordinates are hence a vector with $d$ elements. 
A cell is a {\em neighbor} of another if and only if the coordinates of the two cells differ by exactly one unit in exactly one position.

  Each cell $i$ of the lattice has an integer-valued
state variable $h_{i}$ that represents the floc size at that
cell. We call the units that form the flocs {\em particles}. The
automaton is initialized by placing uniformly and independently at random a total of $m$ particles in the
$n$ cells of the lattice, such that the initial average floc-size is given by $\bar{h} =
\tfrac{m}{n}$. A set of local rules, adapted from those of Almaguer et
al.\ \cite{almaguer}, governs the change of each $h_{i}$
between the current state of the automaton and the next state, in terms
of an {\em aggregation rate} $f$  and a {\em fragmentation rate} $v$.

\begin{description}
\item[Rule 1: if $h_{i} \leq \bar{h}$]{then, with probability $v$, cell $i$
 {\em absorbs} the $\lfloor\bar{h}\rfloor
  - h_{i}$ {\em excess} of each of its neighbors.}
\item[Rule 2: if $h_{i,j} > \bar{h}$]{then, with probability $f \cdot (1 - v)$, the
 cell $i$ {\em absorb} all the particles of all of its
 neighbors (as in Rule 1); otherwise the
 cell {\em transfers} a fraction uniformly at random of its $\lfloor\bar{h}\rfloor
  - h_{i}$ {\em excess} particles, choosing the receptor cell uniformly at random among its
  neighbors.}
\end{description}

First, at each time step, each cell $i$ is processed in a random order. 
Then, according to the Rules 1 and 2, each cell $i$ updates its receptions or transfers according to the values of the state 
variables of the cell itself and its neighbors in the present state. An asynchronous update scheme is performed, 
in such a way that the new state of a cell affects the calculation of states in neighboring cells. 
At the end of the simulation, the floc-size distribution is obtained from the automaton. 

These rules 
govern local fluctuations on the particle exchange among the flocs 
in terms of the relative size of each floc at each cell $i$ with 
respect to the stability {\em threshold}, the average floc-size 
$\bar{h}$. Notice that involves local fluctuations in which mass can be transported by fracture
of large clusters to small ones. However, because of the asymmetry in the absorption 
rates of the neighborhood particles, the mean field description given in \ref{smol} is adequate, 
as argued by the experiments given below.

\begin{figure}[t!]
	\centering
		\includegraphics[width=0.8\textwidth]{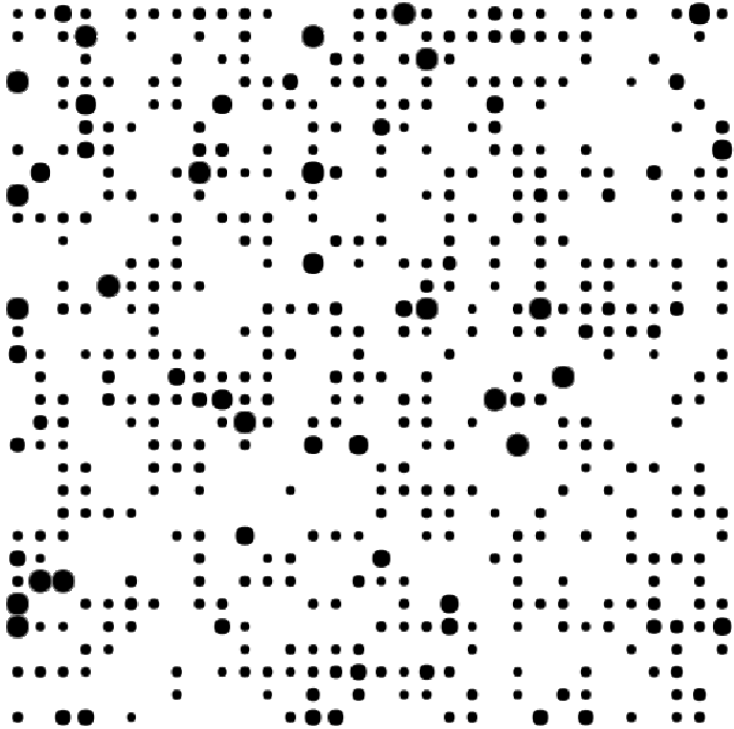}
	\caption{A visualization of a stationary state of a $2$-d
          automaton. The area of each dot is linearly proportional to
          the floc size of the cell.}
	\label{visual}
\end{figure}

A visual representation 
of a typical stationary state of the described $2$-d model is given 
in Figure \ref{visual}.

The parameter $v$ governs floc {\em instability}: when $v \to 1$,
only flocs of sizes below or equal to the average are stable. The
parameter $f$, on the other hand, controls floc {\em stabilization}:
floc growth depends on $f$ when $v < 1$. Therefore the limit $\frac{F}{K} \to 0$ corresponds to $f \to 1$, $v \to 0$ and the limit $\frac{F}{K} \gg 1$
corresponds to $f \to 0$, $v \to 1$. We now show that this microscopic model displays the average behavior predicted by
our Smoluchowski-type model. 
In the absence of fragmentation, the final state should be given by 
a gel state.  
We initialize the automaton starting each cell with a floc of size one. 
There are $n$ total cells. The parameters are fixed to $f = 1$ 
(pure aggregation) and $v = 0.001$ (low fragmentation rate, it cannot be equal to zero). The simulation stops when a singular cell contains a floc of size $n$. 
Ten replicas of the simulation were executed for dimensions from two to five,
from which a stopping criteria for the automaton is defined for each dimension, 
as explained in \cite{benavides}.
In the same previous work \cite{benavides} we explain how to calibrate
the automaton parameters and in this way obtain a physical 
time scale corresponding to the automaton discrete steps. 
In that work the steady state particle distributions of an experimental 
jar test apparatus with agglomeration and velocity is studied by
microscopic images. The parameters are adjusted by statistical tests 
on the similarity between the automaton and experimentally observed 
particle size distributions. In that particular application it turned out
that $10^6$ time steps correspond to $7$ minutes of the jar test experiment.
For the stationary states is predicted a linear
dependency on $\bar{h}$ of the mean and a parabolic  
dependence on $\bar{h}$ of the variance of the cluster size
distribution, like stated by the expressions in (\ref{moments2}).
More interestingly, these expressions describe a range of 
size distributions ranging from those in which small clusters dominate
to gel-like situations in which the particles tend to concentrate in a small
number of very large clusters.

\begin{figure}[t!]
	\centering
		\includegraphics[width=0.7\textwidth]{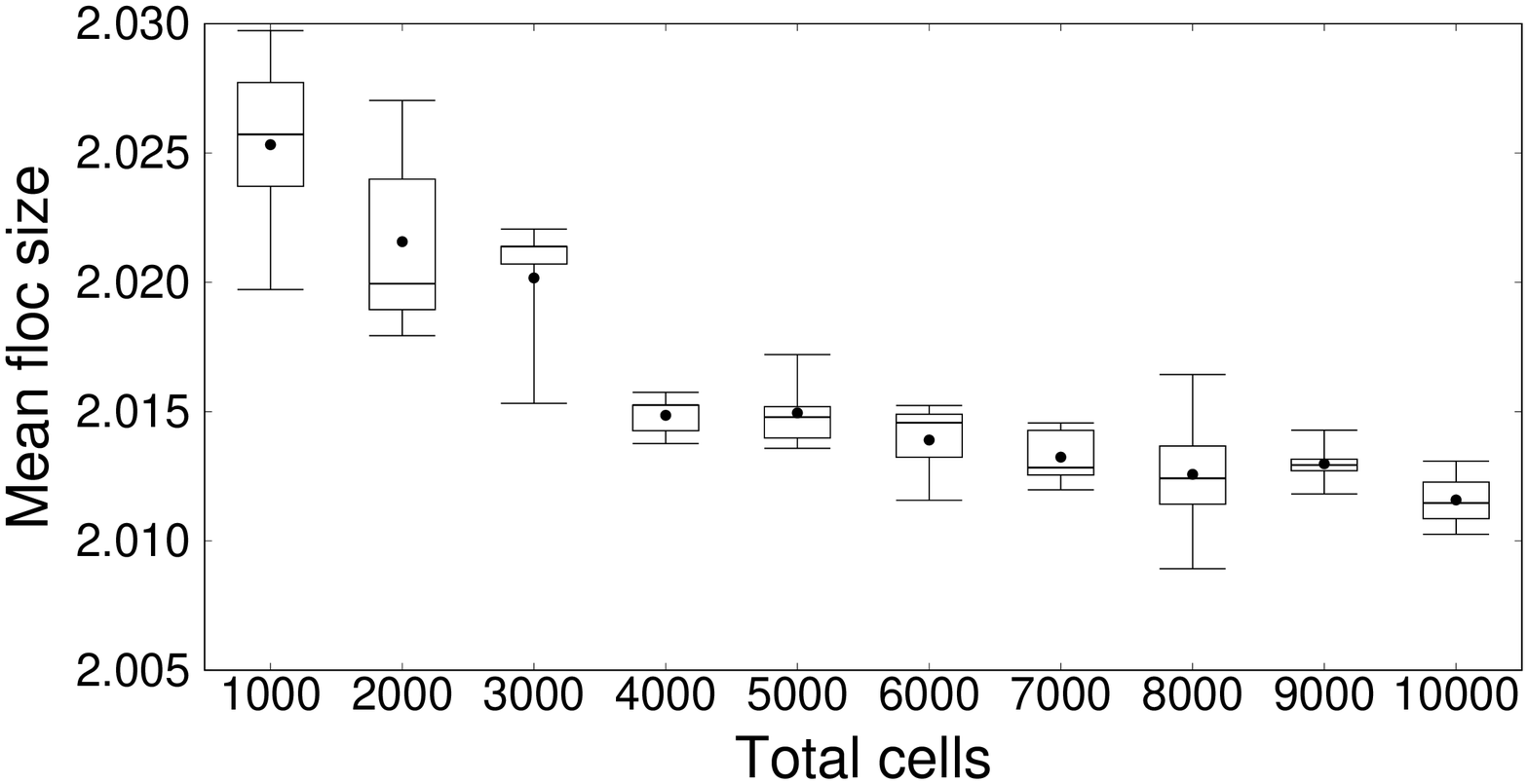}
                \includegraphics[width=0.7\textwidth]{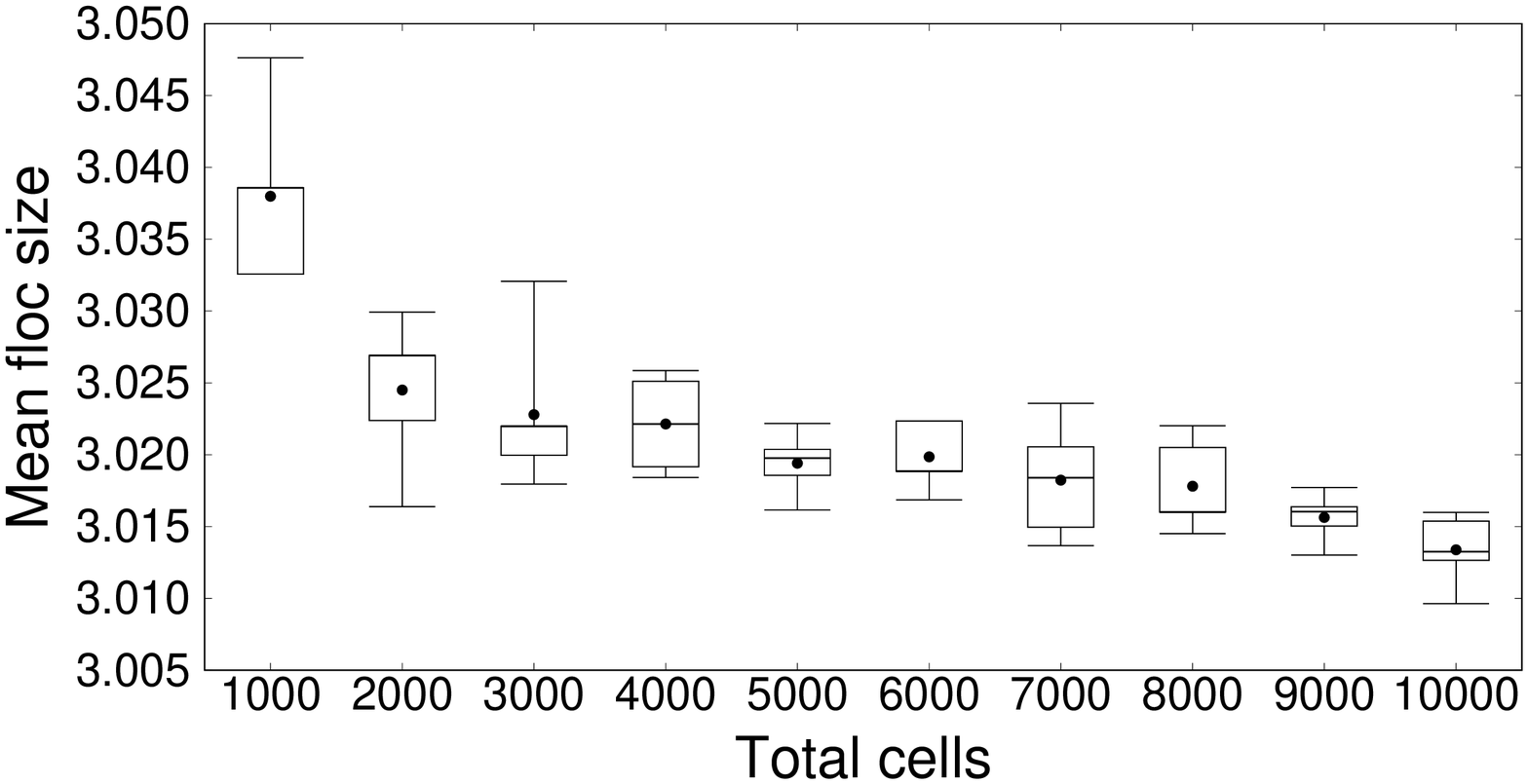} 
                \includegraphics[width=0.7\textwidth]{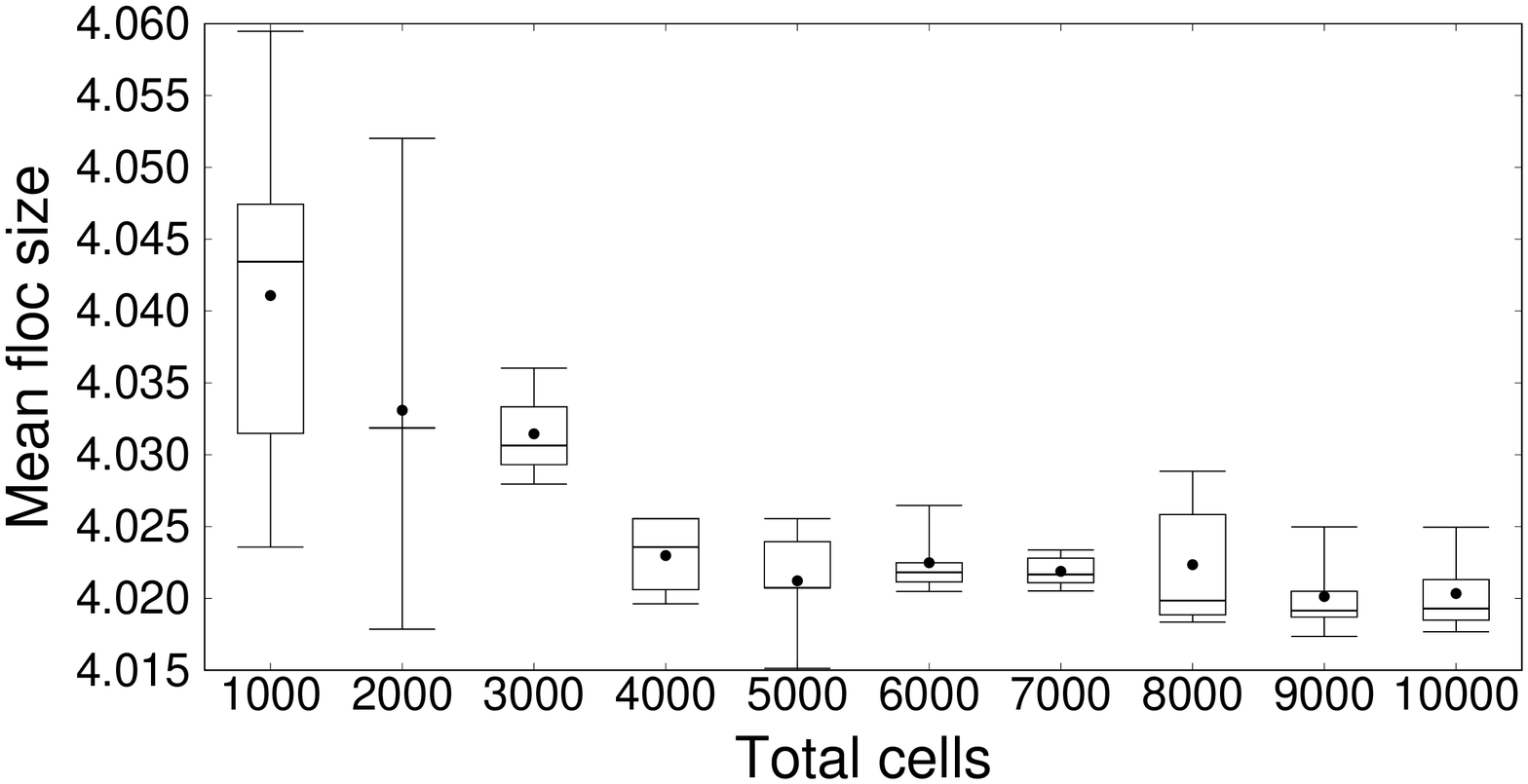}
                \includegraphics[width=0.7\textwidth]{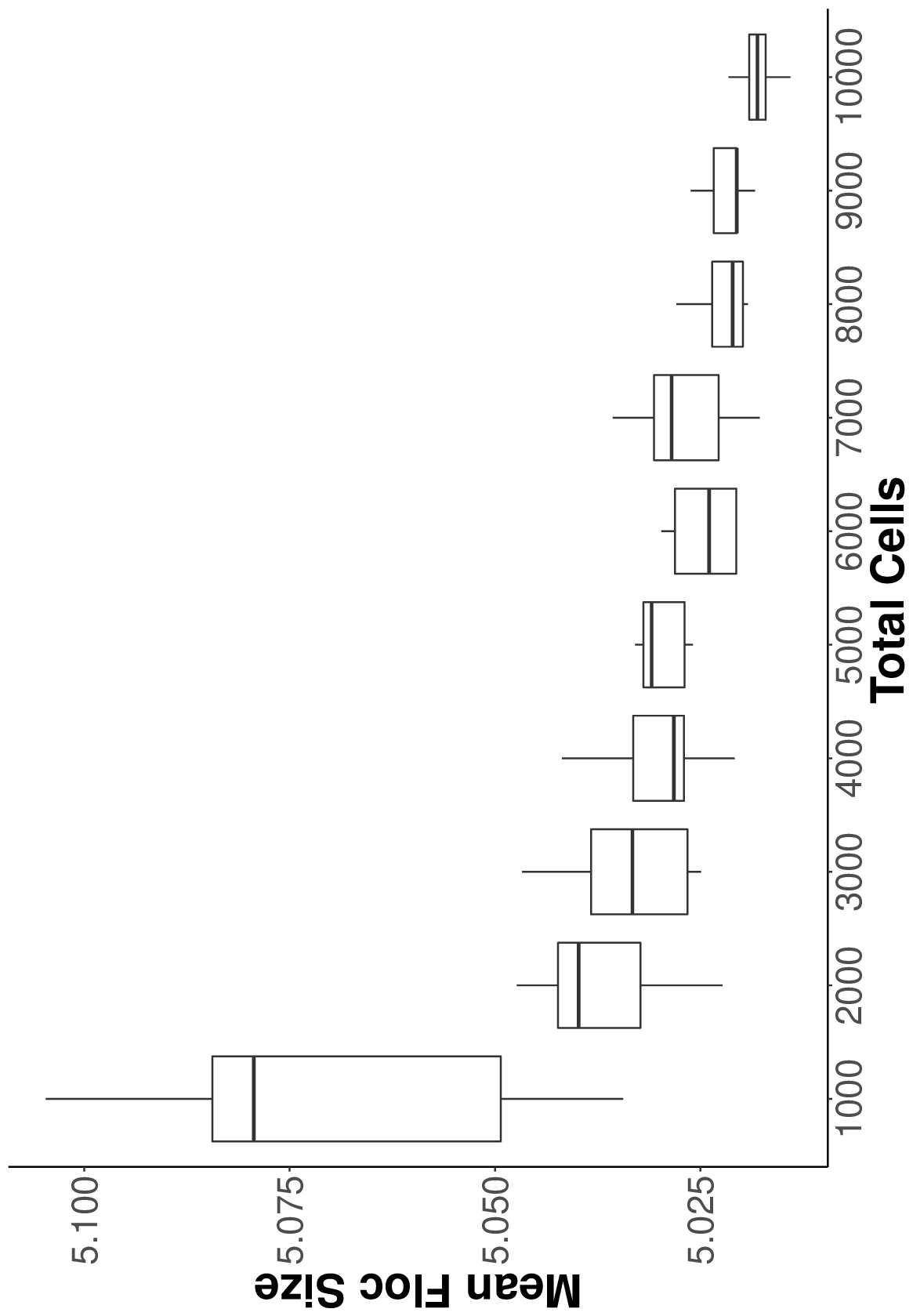}
	\caption{Linear growth in $\bar{h}$ of the first moment of the particle size distribution 
                 of the automaton. The figures correspond (from top to bottom) to 
                 $\bar{h}=2$, $\bar{h}=3$, $\bar{h}=4$, $\bar{h}=5$. The automaton's parameters are $f = 0.1$, $v = 0.9$. The $y$ axis is the 
                 first moment or average floc size and the $x$ axis is the system size.
                 Has can be seen, at large system sizes the average floc sizes become
                 independent of the size but linearly dependent on $\bar{h}$. }
	\label{mean1}
\end{figure}

\begin{figure}[t!]
	\centering
		\includegraphics[width=0.7\textwidth]{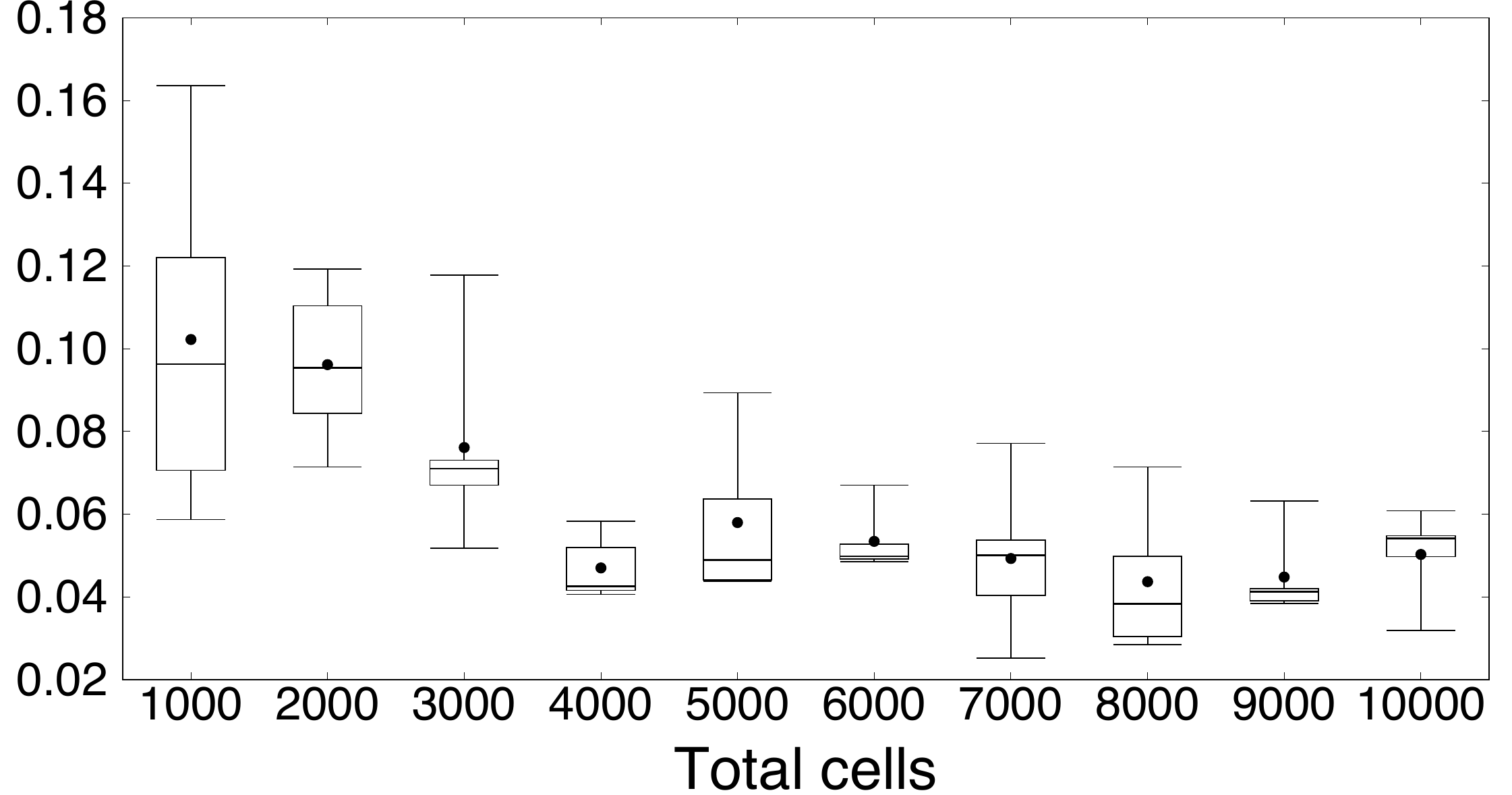}
                \includegraphics[width=0.7\textwidth]{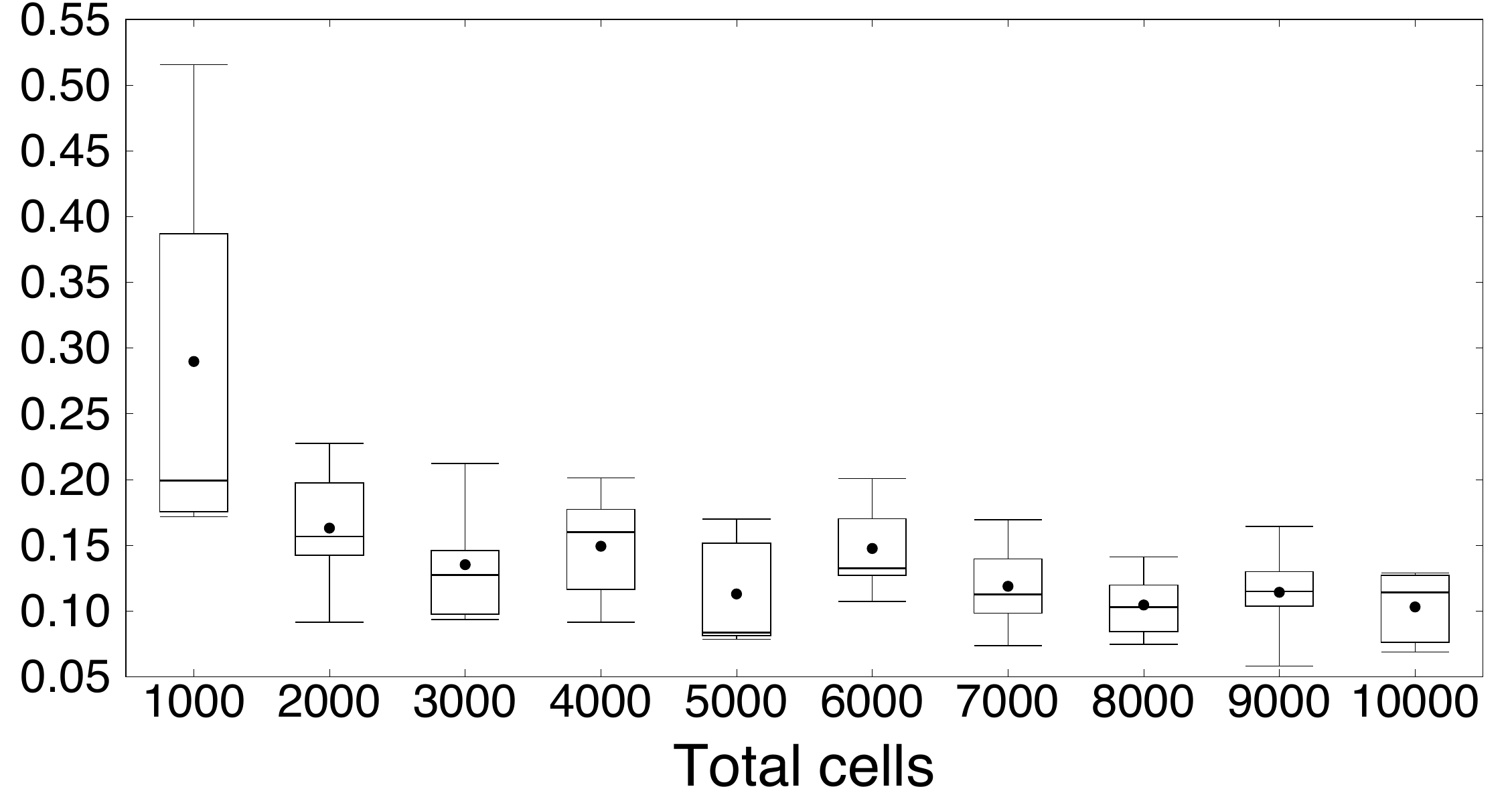} 
                \includegraphics[width=0.7\textwidth]{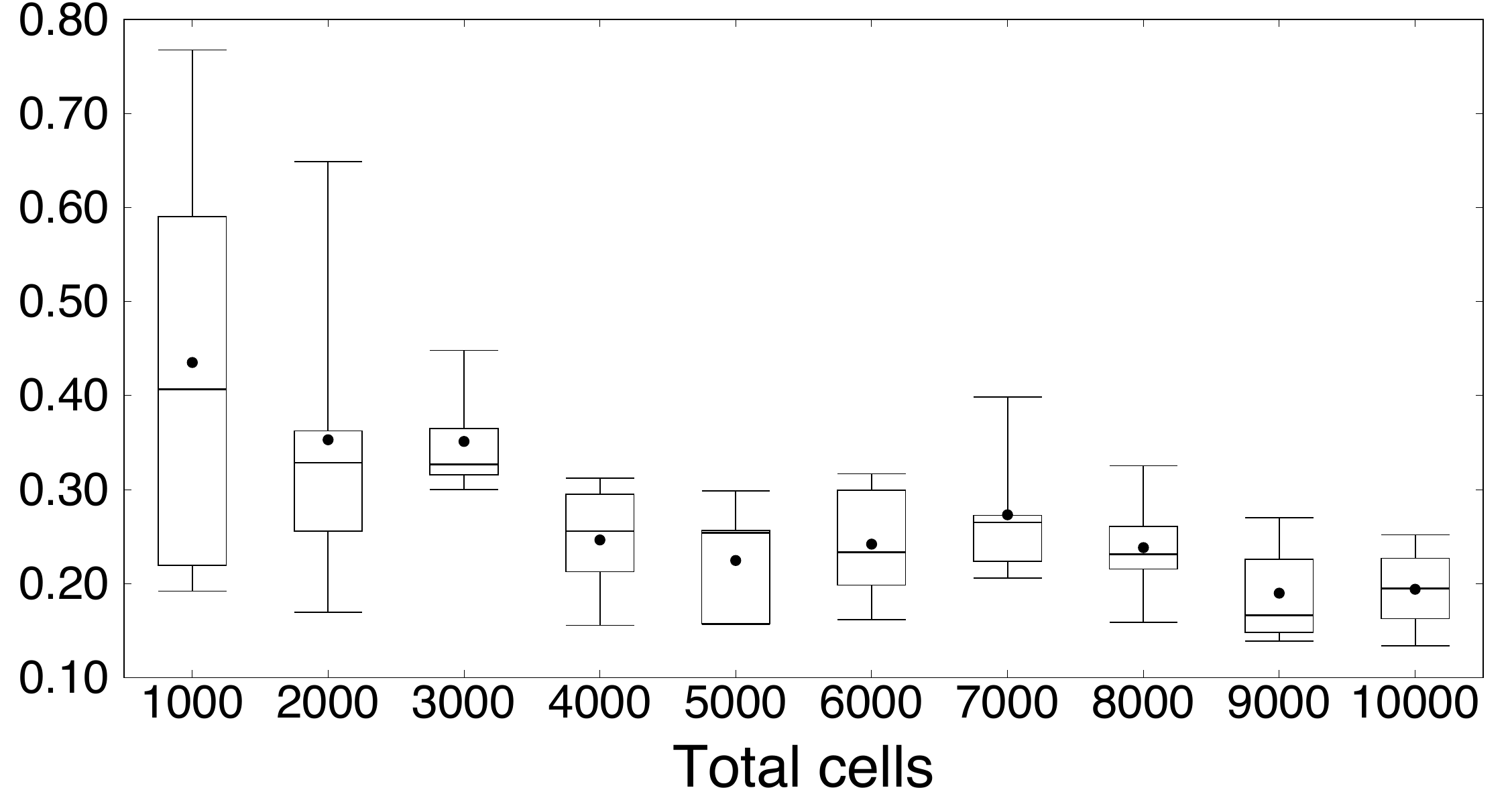}
                \includegraphics[width=0.7\textwidth]{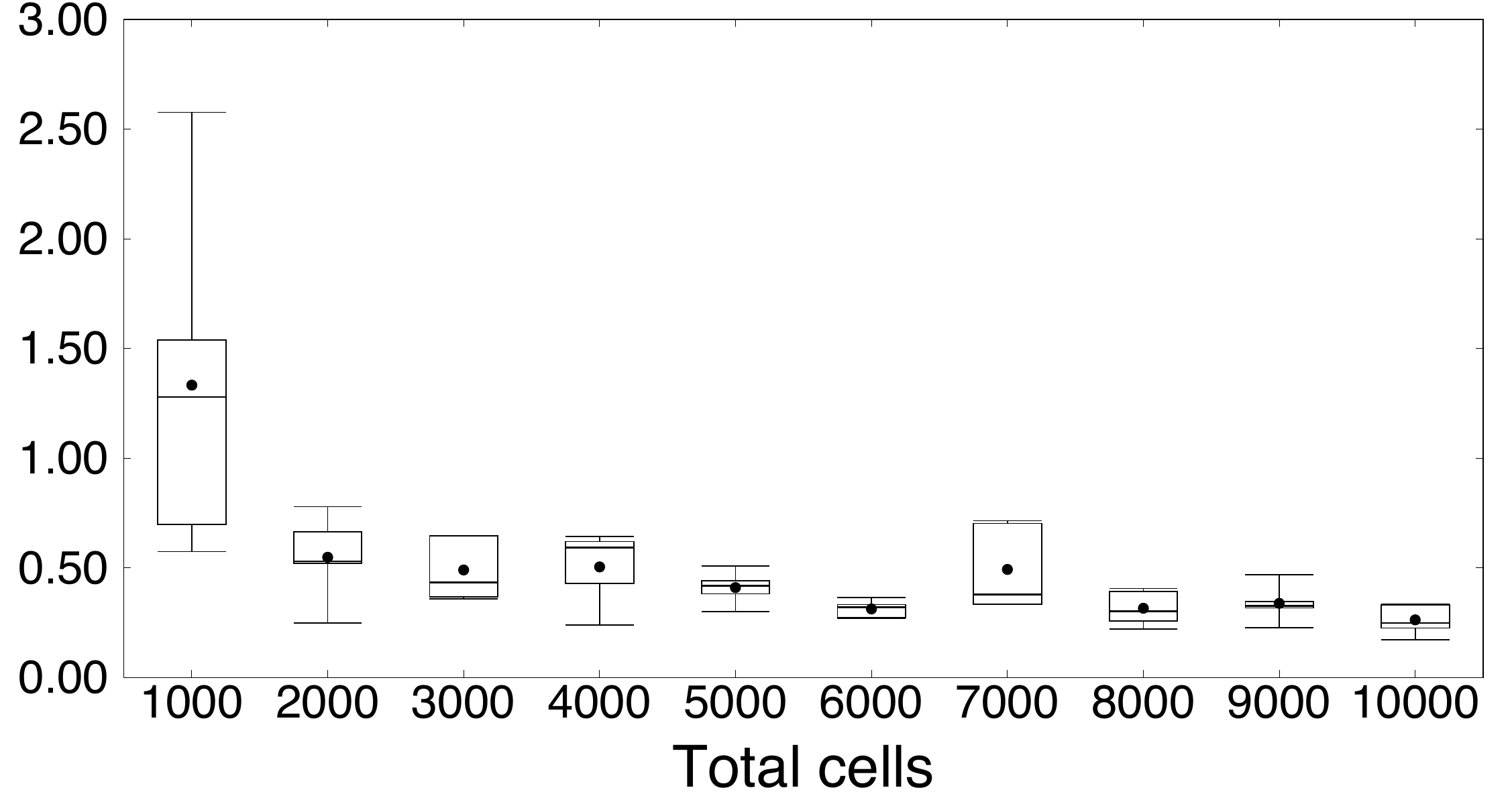}
	\caption{Variance of the same setup as in Figure \ref{mean1}.}
	\label{var1}
\end{figure}

\begin{figure}[t!]
	\centering
                \includegraphics[width=0.7\textwidth]{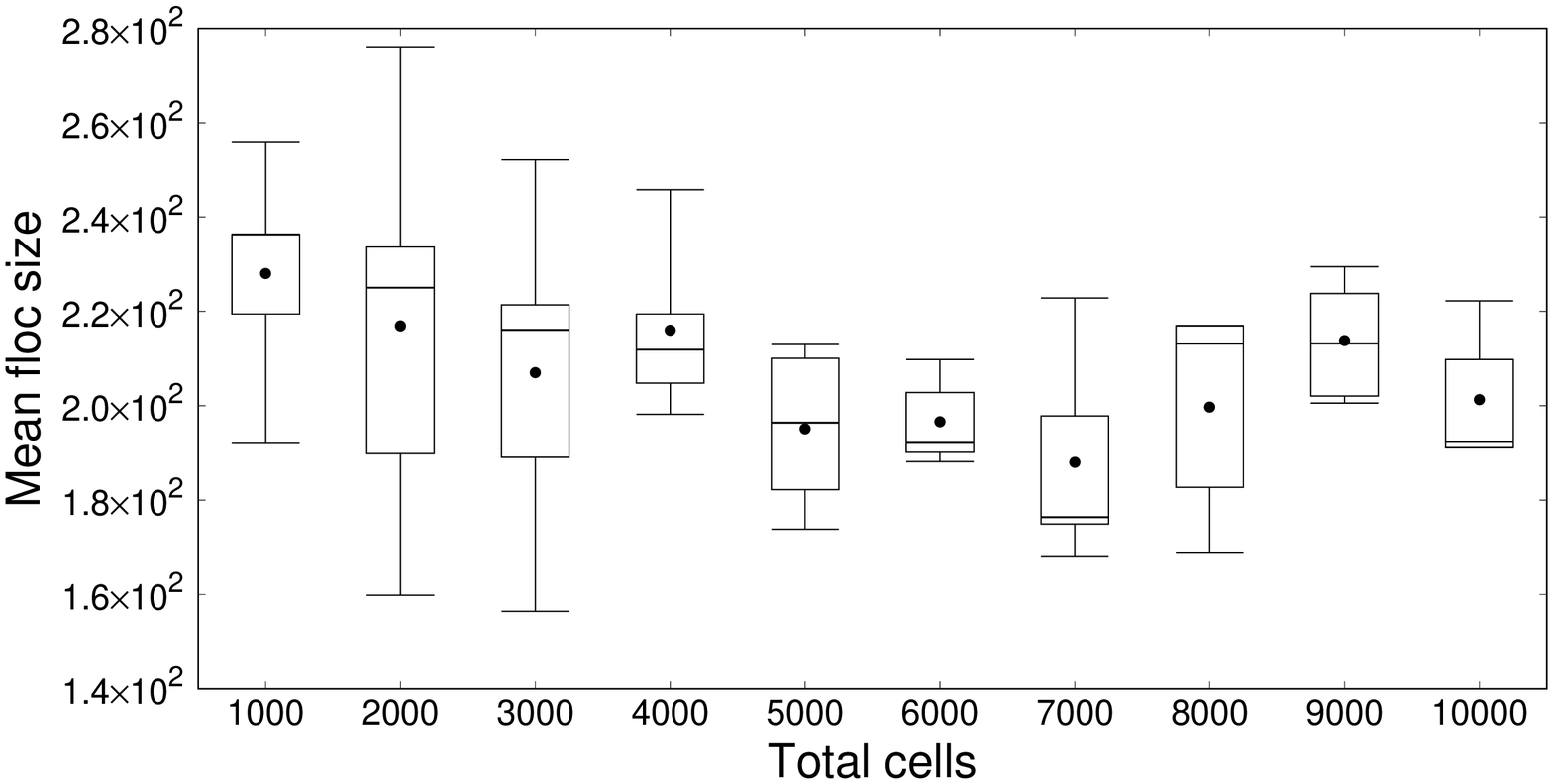} 
                \includegraphics[width=0.7\textwidth]{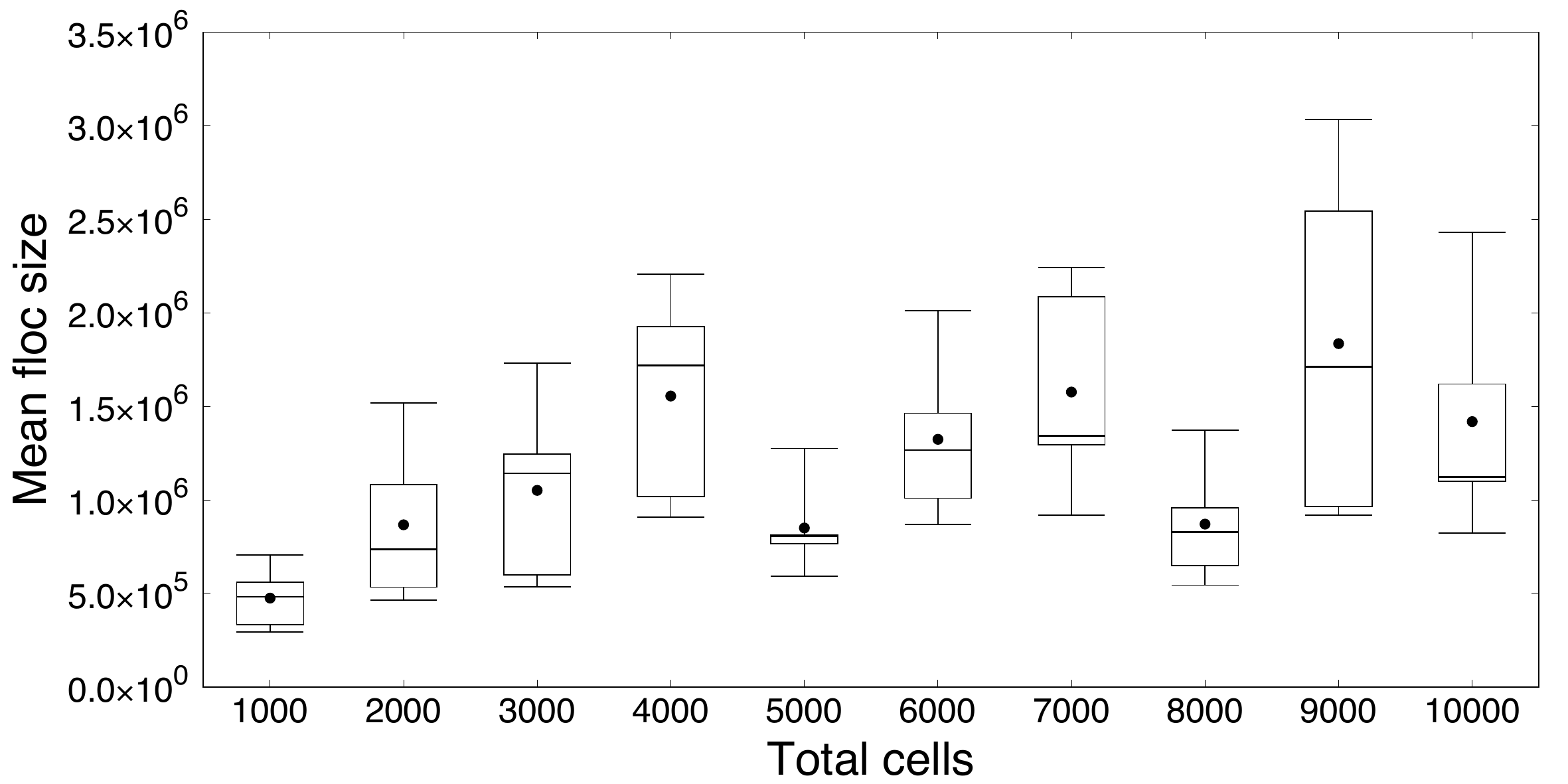} 
	\caption{A stationary state with large mean and variance relative to $\bar{h}=3$ is reported for $f = 0.9$, $v = 0.1$. This is a gel-like state.}
	\label{mean_var}
\end{figure}

In Figure \ref{mean1} we show the average floc size at four different values of $\bar{h}$ with the automaton settings $f = 0.1$, $v = 0.9$.
Is clear the linear dependence with $\bar{h}$ predicted by (\ref{moments2}). 
The variance under the same setup is reported in 
Figure \ref{var1}, which again is consistent with the behavior predicted by equation (\ref{moments2}).
A stationary state with large values of mean and variance relative to $\bar{h}$ is on the other hand reported in 
Figure \ref{mean_var}. In this case the parameters of the automaton are given by
$f = 0.9$, $v = 0.1$ which is consistent with the predictions for $\frac{F}{K} \to 0$. In all cases the reported figures correspond to
$10$ repetitions for each setup, with $10000$ time steps. We have observed that for this number of time steps the particle size
distribution is approximately stationary.

\begin{figure}[t!]
	\centering
		\includegraphics[width=0.7\textwidth]{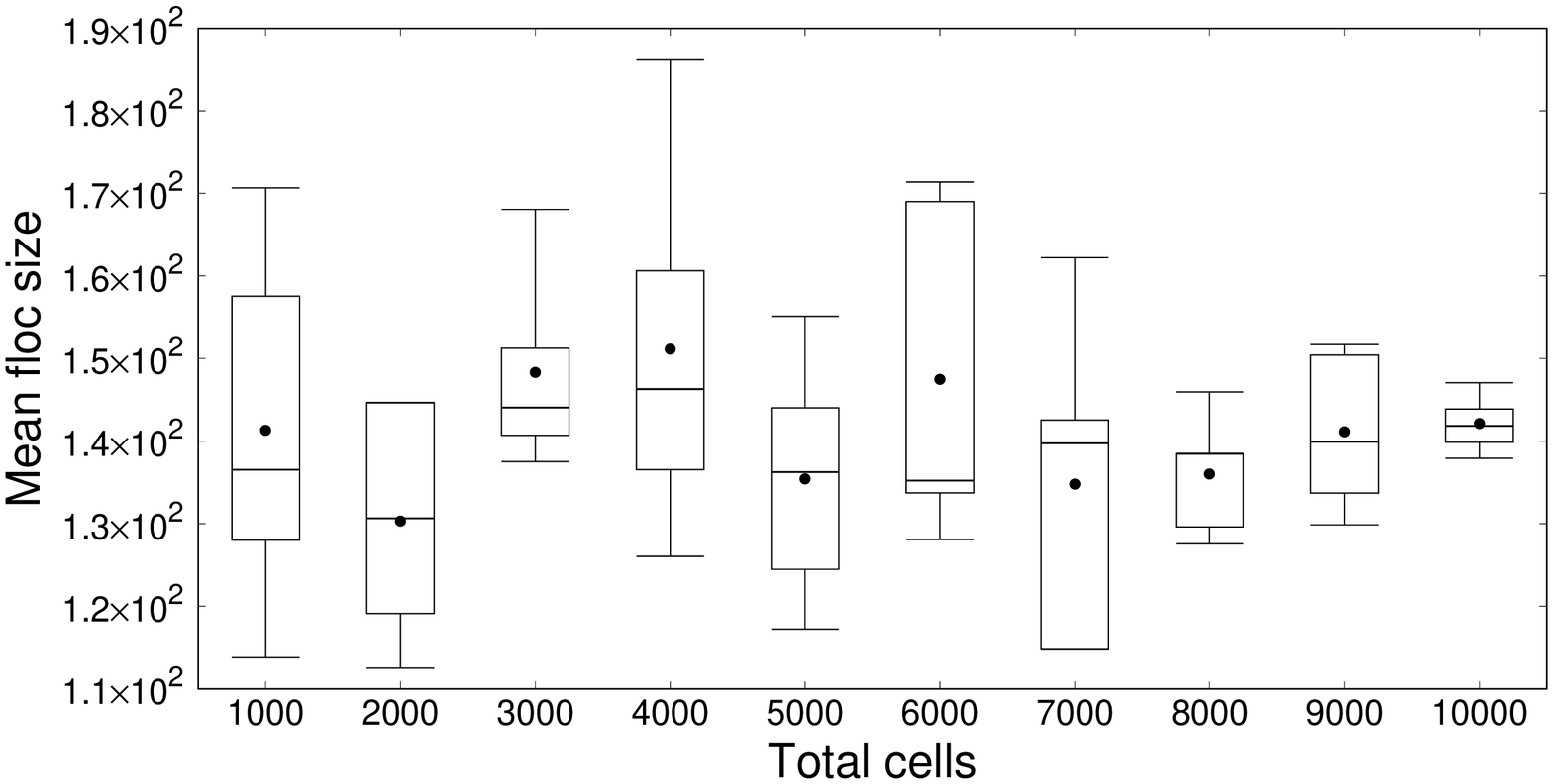}
                \includegraphics[width=0.7\textwidth]{mean_h3c.eps} 
                \includegraphics[width=0.7\textwidth]{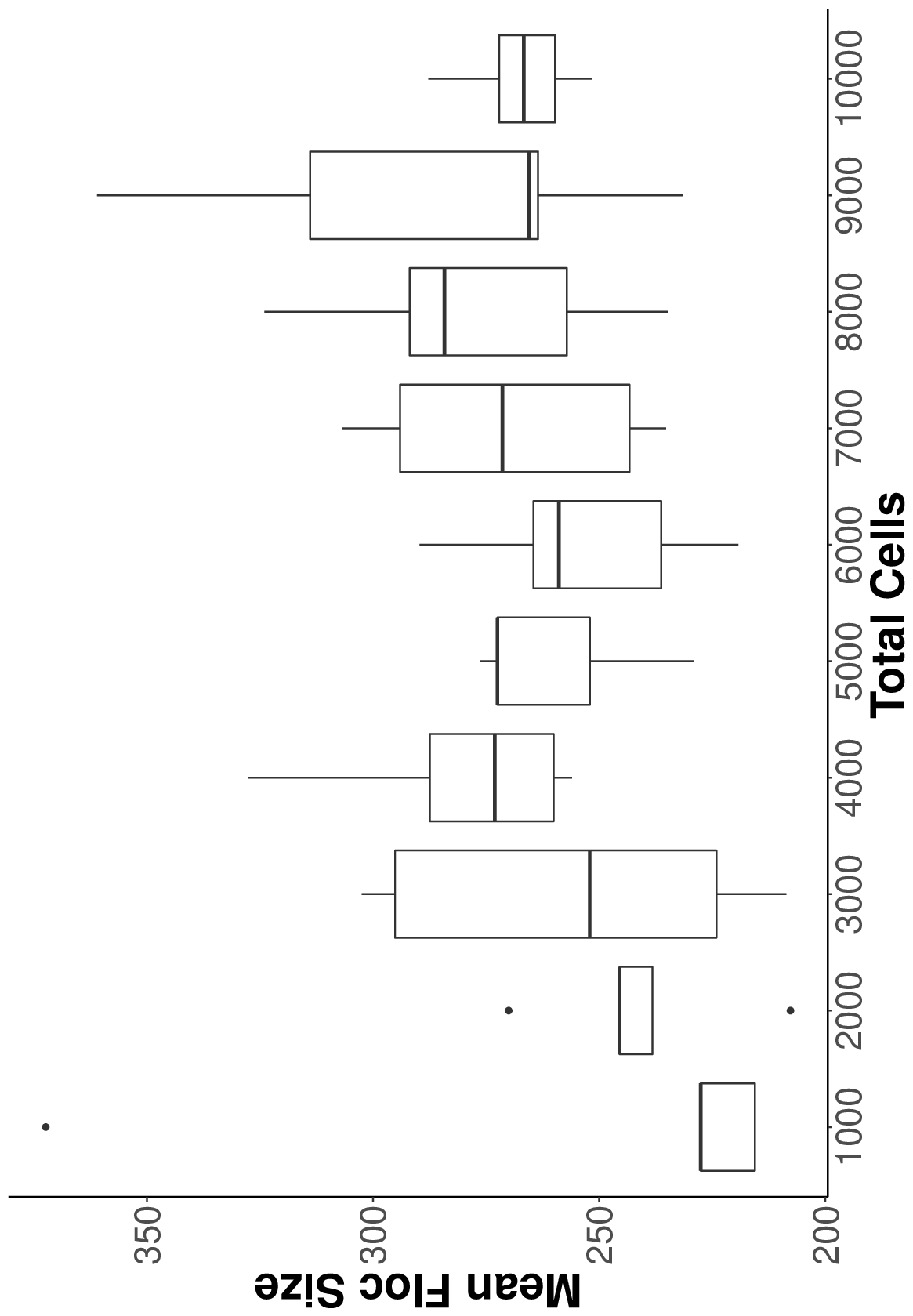}
                \includegraphics[width=0.7\textwidth]{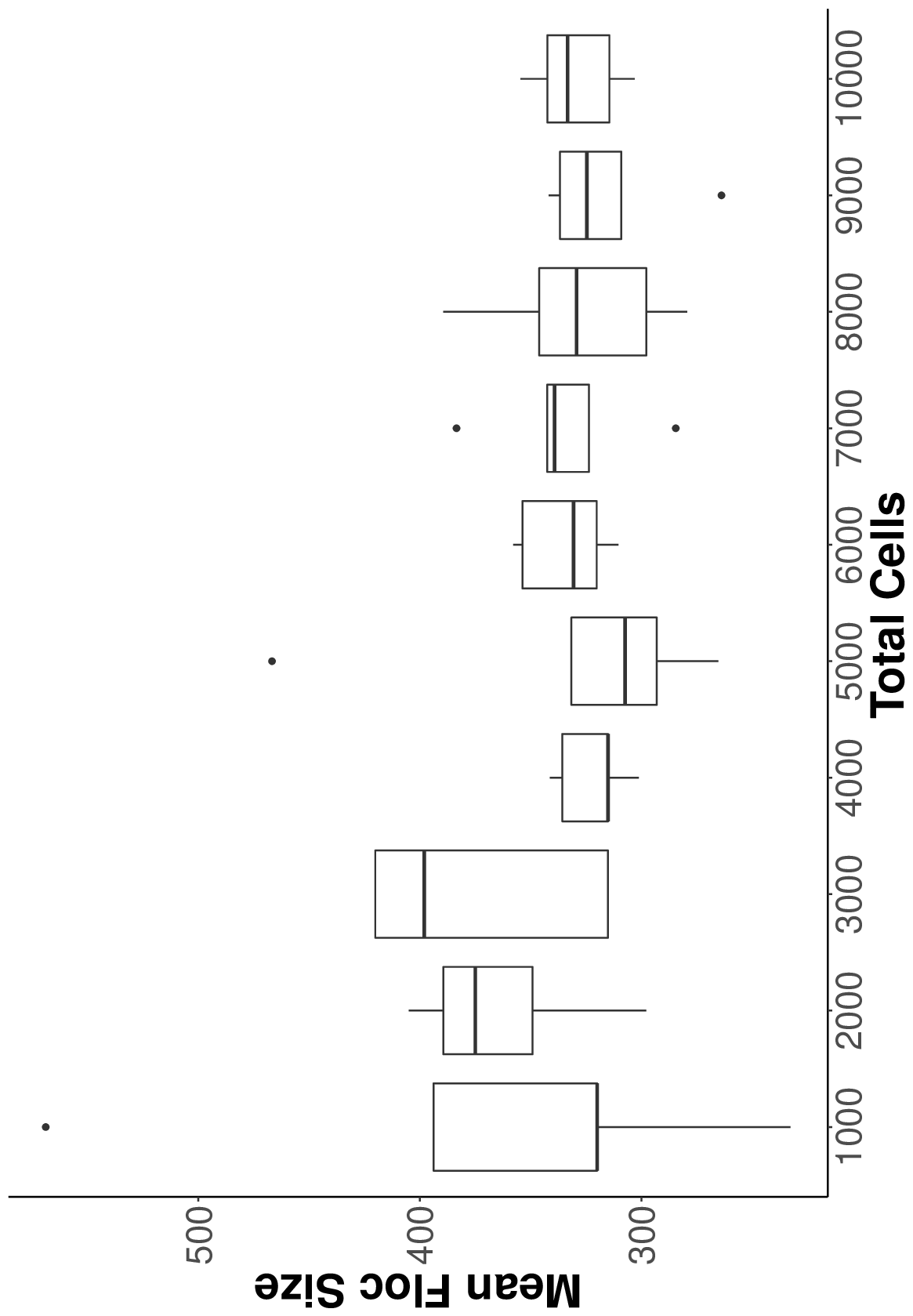}
	\caption{Linear growth in $\bar{h}$ of the mean floc size in a
	stationary gel-like state of the automaton. The figures correspond (from top to bottom) to $\bar{h}=2$, $\bar{h}=3$, $\bar{h}=4$, $\bar{h}=5$. The automaton's parameters are $f = 0.9$, $v = 0.9$.}
	\label{mean2}
\end{figure}

\begin{figure}[t!]
	\centering
		\includegraphics[width=0.7\textwidth]{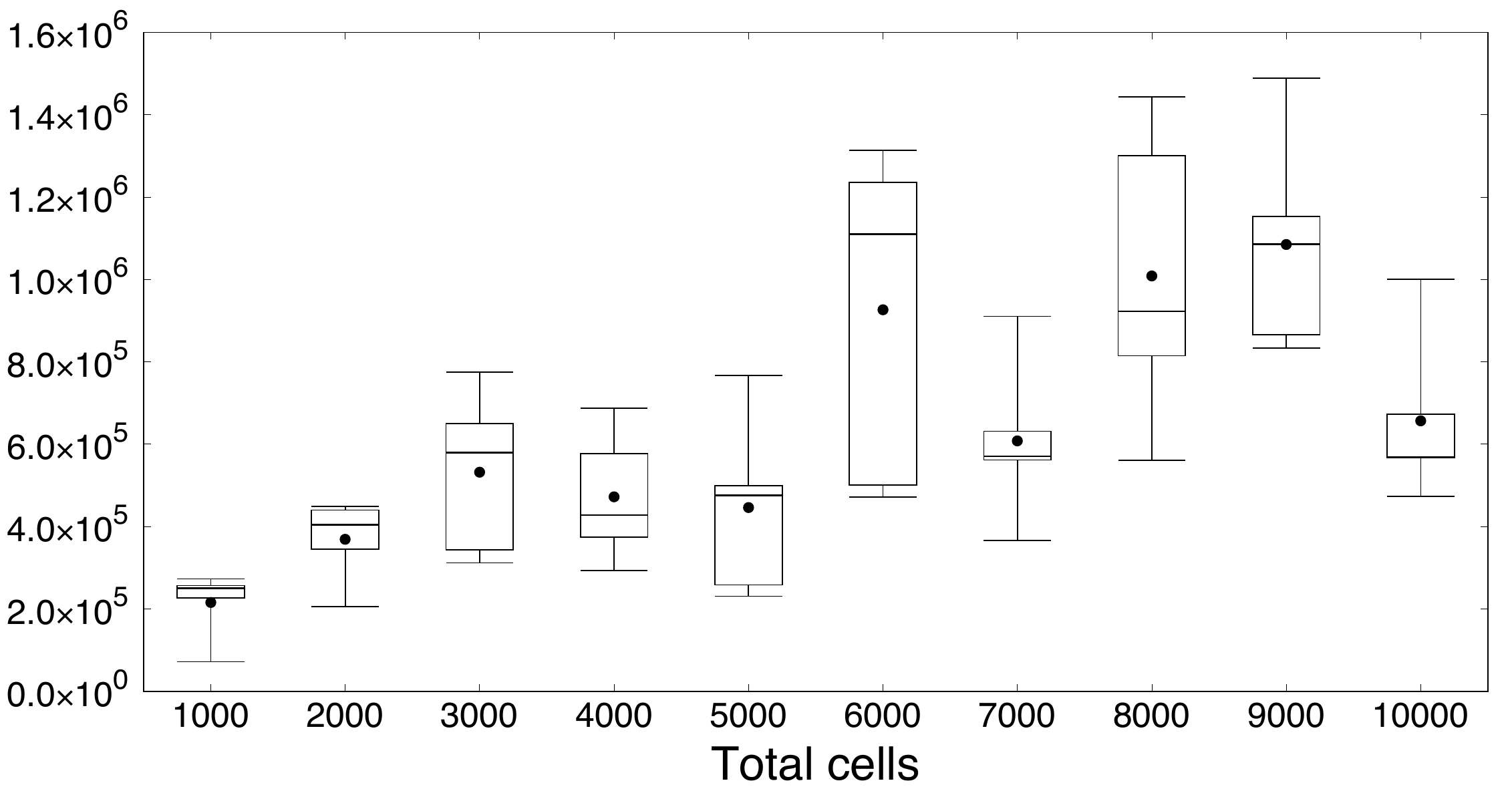}
                \includegraphics[width=0.7\textwidth]{variance_h3c.pdf} 
                \includegraphics[width=0.7\textwidth]{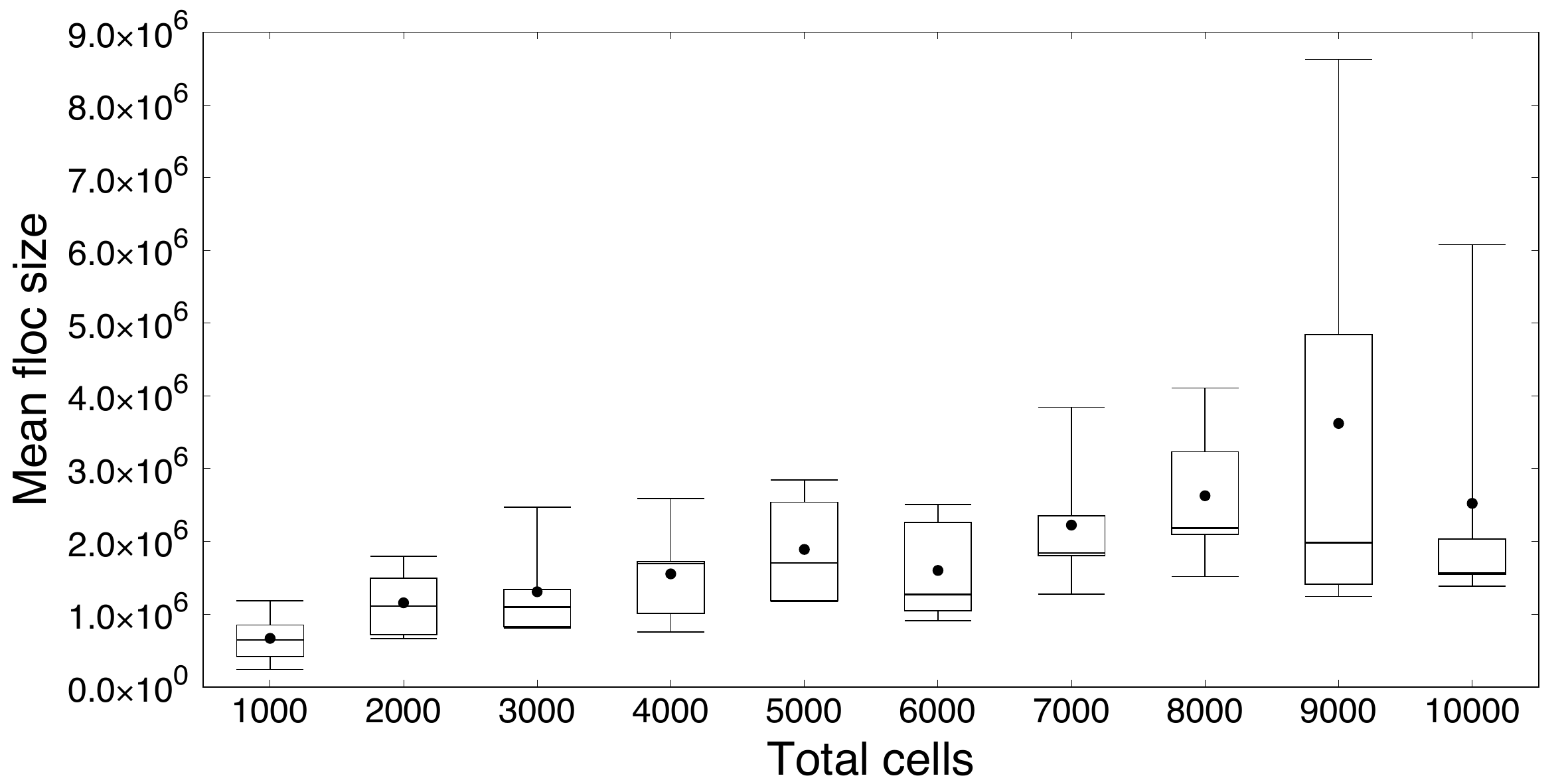}
                \includegraphics[width=0.7\textwidth]{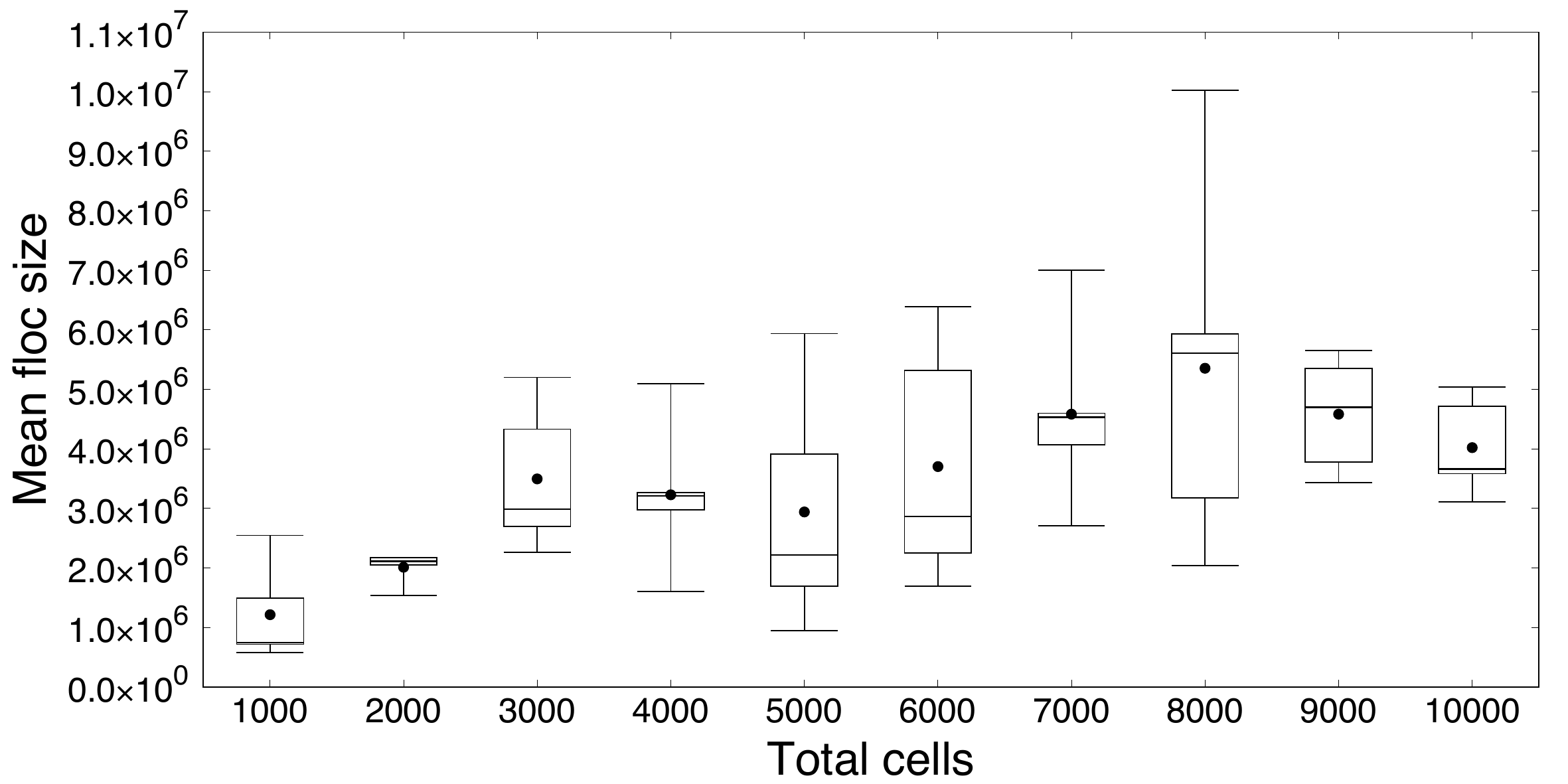}
	\caption{Variance of the same setup as in Figure \ref{mean2}.}
	\label{var2}
\end{figure}

\section{Conclusion} \label{conc}
A new Smoluchowski-type aggregation-fragmentation model has been introduced for
which is possible to analytically derive important statistical features.
To our knowledge, this is the first Smoluchowski-type
model with variable sized fragmentation and instability
that is amenable for analytical treatment. For the constant kernels case,
a microscopic counterpart has been constructed in terms of a cellular automaton.
The simulations are in agreement with the predictions derived from the Smoluchowski-type model.
The framework introduced in this work is relevant to applications in which 
populations of clusters that can aggregate become unstable at 
sizes above some threshold. For instance, there is experimental evidence that this
might be the case in the removal of heavy metals from wastewater through coagulation and
agitation \cite{almaguer}. Other possible application of the model is in social networks, 
where intuitively is plausible that large communities above some size are more prone to fracture into smaller communities.

\section*{Acknowledgement} 

This work was partially supported by supported by UANL-PAICYT under grant IT1807-21
and by CONACYT under grant CB-167651.

\end{document}